\documentclass[aps,prd,preprint]{revtex4}

\usepackage{graphicx}
\usepackage{amsmath}

\begin{document}

\title{Solution of the Stochastic Langevin Equations for Clustering of Particles in Random Flows in Terms of Wiener Path Integral}
\author{M. Chaichian$^{1,2}$}\email{masud.chaichian@helsinki.fi}
\author{A. Tureanu$^{1,2}$}\email{anca.tureanu@helsinki.fi}
\author{A. Zahabi$^1$}\email{seyedali.zahabi@helsinki.fi}
\affiliation{ $^1$Department of Physics, University of Helsinki,
P.O. Box 64, FIN-00014 Helsinki, Finland}
\affiliation{ $^2$ Helsinki Institute of Physics, P.O. Box 64, FIN-00014 Helsinki, Finland}
\date{\today}

\begin{abstract}
\noindent

We propose to
take advantage of using the Wiener path integrals as
the formal solution for the joint probability densities of coupled
Langevin equations describing particles suspended in a fluid under
the effect of viscous and random forces. Our obtained formal
solution, giving the expression for the Lyapunov exponent, \emph{i})
will provide the description of all the features and the behaviour
of such a system, e.g. the aggregation phenomenon recently studied
in the literature using appropriate approximations, \emph{ii}) can
be used to determine the occurrence and the nature of the
aggregation - non-aggregation phase transition which we have shown
for the one-dimensional case and \emph{iii}) allows the use of a
variety of approximative methods appropriate for the physical
conditions of the problem such as instanton solutions in the WKB
approximation in the aggregation phase for the one-dimensional case
as presented in this paper. The use of instanton approximation gives
the same result for the Lyapunov exponent in the aggregation phase,
previously obtained by other authors using a different approximative
method. The case of non-aggregation is also considered in a certain
approximation using the general path integral expression for the
one-dimensional case.
\end{abstract}

\maketitle

\section{Introduction}

Recently, there has been considerable interest in the study of the behaviour of particles in media taking into account the effect of random forces. Studies in this direction can provide a better understanding of the behaviour of particles in turbulent flow. The features and behaviour of turbulent flow are under continuous intense investigations. As one of the main features, the clustering of particles into regions of high density has been studied extensively on both experimental and theoretical sides
\cite{ref:1.1}-\cite{ref:1.4}.
Particles suspended in a turbulent fluid form cluster structures as a result of the competition between the diffusive random forces and the aggregative viscous ones. However, the conditions for such a behaviour are not fully understood, the mechanisms which contribute to the formation of clusters have been studied in \cite{ref:1.5}-\cite{ref:1.8} (see also \cite{ref:1.9}).
An extreme form of clustering of particles, known as the "aggregation phenomenon", which is not well-understood, has been studied recently by means of theoretical modeling and numerical simulations \cite{ref:1.10}-\cite{ref:1.16}. Other phenomenological models for cluster aggregation, inspired by Kolmogorov's theory \cite{ref:1.17}, may be also studied along similar lines \cite{ref:1.18}. The aggregation of particles can be defined as the coalescence of different particles paths with very close positions and velocities in a fluid subjected to random forces fluctuating in space and time, the particles being affected by viscous forces proportional to their velocities. The first theoretical analysis and numerical simulations for the aggregation of suspended particles in a one-dimensional random fluid were carried out in \cite{ref:1.10}. The result of this study shows a phase transition between the non-aggregate and aggregate phases. Motivated by this result, recent investigations on the aggregation of particles in two- and three-dimensional random fluids were performed in \cite{ref:1.11}-\cite{ref:1.13}. By introducing a model for the motion of point-like non-interacting particles in a three dimensional random fluid, the equations of motion for such particles, which are under the influence of a viscous force beside the random force were also derived in \cite{ref:1.13}. Then by linearizing the equations of motion, two coupled Langevin equations which describe the evolution of the separation of positions and velocities of two nearby particles were obtained. These two coupled Langevin equations describe the aggregation of particles and thus the system of coupled Langevin equations should be solved for calculating the Lyapunov exponent \cite{ref:1.19}, which is equal to the expectation value of one of the variables in the Langevin equations. For this purpose, we apply the Wiener path integral formalism for solving the system of two coupled Langevin equations, describing the aggregation phenomenon. At first we introduce a method for writing the solution of Langevin equations in terms of the Wiener path integral, which has been thoroughly studied in the literature over the years \cite{ref:1.20}-\cite{ref:1.24}. Then, by generalizing the procedure, we obtain the solution of the system of $\emph{N}$ coupled Langevin equations in terms of the Wiener path integral. The Lyapunov exponent as an indicator of aggregation can also be written in terms of path integral. The Wiener path integral formalism provides an exact solution to the aggregation problem. The obtained exact solution in terms of the path integral is presented in a closed analytical form and its actual evaluation can be performed by means of a variety of different approximation methods suitable to the specific physical conditions of the system. We discuss some approximative methods for such Wiener path integrals in one-dimensional case.

\section{System of coupled Langevin equations describing the aggregation phenomenon }
In this section we briefly review the results of recent studies \cite{ref:1.13} on the aggregation phenomenon in three dimensions.
Particles suspended in a turbulent fluid can be modeled by the spherically massive particles which are moving in a random velocity field with specific properties such as isotropic, homogeneous, and stationary statistics. For simplicity, it can be assumed that there is no interaction between the particles themselves as well as between the particles and the fluid. Also we can neglect the inertia of the displaced fluid. By these assumptions, one can consider a large number of suspended particles with random initial positions in the fluid and zero velocities. The behaviour of the trajectories of particles is dictated by the effect of random as well as viscous forces and the motion of such particles in a random fluid is diffusive. Therefore, the inhomogeneities in density tend to get reduced, while the viscous forces cause the aggregation of particles and eventually the competition between diffusive random forces and viscous forces leads to a phase transition between path coalescence and path non-coalescence phases.
The equations of motion which describe the suspended particles' trajectories are
\begin{equation}
    \dot{\textbf{r}} = \frac{\textbf{p}}{m} , \hspace{1cm} \dot{\textbf{p}} = -\gamma [\textbf{p} - m\textbf{u} (\textbf{r},t)],
\end{equation}
where $\gamma$ characterizes the strength of the viscous damping and $\textbf{u} (\textbf{r},t)$ is the random velocity field.
The aggregation of particles can be studied by considering two nearby trajectories with spatial separation $\delta \textbf{r}$ and momenta difference $\delta \textbf{p}$. The linearized version of the equation of motion can be derived as
\begin{equation}
    \delta \dot{ \textbf{r}} = \frac{\delta \textbf{p}}{m} , \hspace{1cm} \delta \dot{ \textbf{p}} = -\gamma \delta \textbf{p} + \textbf{F} (t) \delta \textbf{r},
\end{equation}
with the matrix elements of \textbf{F} as
\begin{equation}
    F_{ij} = \frac{\partial f_i}{\partial r_j} (\textbf{r} (t),t) = m \gamma \frac{\partial u_i}{\partial r_j} (\textbf{r} (t),t).
\end{equation}
With the parametrization of the linearized equations of motion as
\begin{equation}
    \delta \textbf{r} = X\textbf{n}_1,  \hspace{1cm}    \delta \textbf{p} = X (Y_1\mathbf{n}_1 + Y_2\mathbf{n}_2),
\end{equation}
where $X$ is a scale factor showing the aggregation occurs if it decrease with probability in the limit $t\rightarrow\infty$. $\textbf{n}_1$ and $\textbf{n}_2$ are time-dependent orthogonal unit vectors (in three dimensions the third unit vector is defined as $\textbf{n}_3 = \textbf{n}_1 \wedge \textbf{n}_2$).
By differentiating eq. (4), substituting the result in eq. (2) and finally projecting the equations onto the unit vectors one obtains the equations of motion for the variables $X$ and $Y_i$:
\begin{eqnarray}
    \dot{X} &=& \frac{1}{m}Y_1 X\nonumber\\
    \dot{Y}_1  &=& - \gamma Y_1 + \frac{1}{m}(Y_2^2 - Y_1^2) + F^\prime_{11} (t),\nonumber\\
    \dot{Y}_2  &=& - \gamma Y_2 - \frac{2}{m}Y_1Y_2 + F^\prime_{21} (t),
\end{eqnarray}
where $\textbf{n}_i(t).\textbf{F}(t)\textbf{n}_j(t) = F^\prime_{ij} (t)$.
It has been argued \cite{ref:1.13}, that the maximal Lyapunov exponent $\lambda_1$ (hereafter called simply Lyapunov exponent)
in the limit $t\rightarrow\infty$ is given by
\begin{equation}
    \lambda_1 = \langle \frac{d\ln X}{dt} \rangle = \frac{1}{m} \langle Y_1 \rangle,
\end{equation}
which we used the first equation in (5).

We recall the general definition of the Lyapunov exponent as a quantity that characterizes the rate of separation of infinitesimally close trajectories. Quantitatively, the separation $\delta Z (t)$ of two trajectories in phase space with initial separation $\delta Z_0$ is given by the formula
\begin{equation}
    \mid\delta Z (t)\mid  \approx  \mid\delta Z_0\mid \exp{(\lambda t)} ,
\end{equation}
where $\lambda$ is the Lyapunov exponent. The separation diverges with time when $\lambda > 0$ and aggregation occurs for the case $\lambda < 0$.

In the limit where the correlation time $\tau^\prime$ of the random force is small and the random force itself is also sufficiently weak, the coupled equations of motion can be approximated by a system of two coupled Langevin equations (for details of derivations and physical explanations see part III of \cite{ref:1.13}):
\begin{eqnarray}
    dY_1  &=& \left[-\gamma Y_1 + \frac{1}{m} (Y_2^2-Y_1^2)\right]dt + d\zeta_1,\nonumber\\
    dY_2 &=& \left[-\gamma Y_2 + \frac{D_{31}}{Y_2} - \frac{2}{m} Y_1Y_2\right]dt + d\zeta_2,
\end{eqnarray}
with the noise properties
\begin{equation}
    \langle d\zeta_i \rangle = 0,\hspace{2cm} \langle d\zeta_i d\zeta_j \rangle = 2D_{ij} dt.
\end{equation}
The diffusion constants $D_{ij}$ is defined as
\begin{equation}
    D_{ij} = \frac{1}{2}\int_{-\infty}^{+\infty}\, dt \  \! \langle F'_{i1} (t)F'_{j1} (0) \rangle.
\end{equation}
Consequently, by the change of variables
\begin{eqnarray}
dt' = \gamma dt,\hspace{2cm} x_i = \sqrt{\frac{\gamma}{D_i}} Y_i,\hspace{2cm} dw_i =  \sqrt{\frac{\gamma}{D_i}} d\zeta_i,
\end{eqnarray}
where $D_1 \equiv D_{11},\hspace{0.2cm} D_2 \equiv D_{21} \equiv D_{31}$, the final coupled Langevin equations read as
\begin{eqnarray}
    dx_1  &=& \left[-x_1 + \epsilon(\Gamma x_2^2-x_1^2)\right]dt^\prime + dw_1,\nonumber\\
    dx_2 &=& \left[-x_2 + x_2^{-1} -2\epsilon x_1 x_2\right]dt^\prime + dw_2,
\end{eqnarray}
where $\epsilon = D_1^{1/2}/m\gamma^{3/2}$ is a dimensionless measure of the inertia of the particles and $\Gamma \equiv D_2/D_1$ is a measure of the relative intensities of potential and solenoidal components of the velocity field.
Equivalently, the coupled Langevin equations can be written as
\begin{eqnarray}
    \dot{x}_1  &=& \left[-x_1 + \epsilon(\Gamma x_2^2-x_1^2)\right] + \dot{w}_1,\nonumber\\
    \dot{x}_2 &=& \left[-x_2 + x_2^{-1} -2\epsilon x_1 x_2\right] + \dot{w}_2,
\end{eqnarray}
with
\begin{equation}
    \langle\dot{w_i}\rangle = 0, \hspace{1cm} \langle\dot{w_i} \dot{w_j}\rangle = 2 \frac{\delta_{ij}}{dt^\prime},\hspace{2cm} i,j = 1,2.
\end{equation}
This system of stochastic differential equations should be solved and the derived solution, which represents the probability density, can be used to determine the Lyapunov exponent $\lambda_1$ \cite{ref:1.19}, \cite{ref:1.13}, as an indicator of the aggregation phenomenon:
\begin{equation}
    \lambda_1 = \gamma \epsilon \, \langle x_1 \rangle.
\end{equation}
When the Lyapunov exponent $\lambda_1$ is negative the aggregation phenomenon occurs \cite{ref:1.13}. A positive Lyapunov exponent is an indication that the system is chaotic.

The system of two coupled Langevin equations as a special kind of stochastic differential equations has an exact solution in terms of the Wiener path integral. In the following, we introduce the Wiener path integral as the solution of stochastic differential equations for the special case of the Langevin equations. Subsequently, we generalize the Wiener path integral formalism for the system of coupled Langevin equations, which describe massless as well as massive Brownian particles in random media. After that we shall be able to present a solution in terms of the Wiener path integral for the coupled Langevin equations describing the aggregation phenomenon.

\section{Solution of the stochastic differential Langevin equations in terms of Wiener path integral}
In this section, we introduce the Wiener path integral method for the solution of stochastic differential equations of special kind, the Langevin equations. As a prototype for such Langevin equations, we can look upon them as describing the Brownian motion in different, coordinate or velocity, spaces. For a comprehensive description of several Brownian particles in general, see \cite{ref:1.25}.

\subsection{Wiener path integral for one Langevin equation}
At first, we consider a Brownian particle in a random medium and write the path integral solution for the transition probability of the particle from one fixed arbitrary initial point to a fixed arbitrary final point. The Wiener path integral method can be used for the analysis of the stochastic equations and consists in determining the statistical properties of their solutions such as probability densities and expectation values.
The microscopic approach to stochastic processes starts from the stochastic Langevin equation.
The Langevin equation for a Brownian particle subject to a general non-stationary and nonlinear external force is
\begin{equation}
    m \ddot{x} + \eta \dot{x} = F + \dot{\Phi},
\end{equation}
where $m$ is the mass of the particle, $\eta$ is the friction coefficient, $F$ is an external force and $\dot{\Phi}$ is a random force. For sufficiently large time intervals $t\gg m/\eta $ we can neglect the mass term, so the Langevin equation which describes the motion of inertialess Brownian particles takes the form
\begin{equation}
    \dot{x} (\tau) + f (x(\tau),\tau) = \dot{\phi} (\tau),
\end{equation}
where
\begin{equation}
    f = \frac{F}{-\eta},\hspace{2cm}   \dot{\phi} = \frac{\dot{\Phi}}{\eta}.
\end{equation}
Performing a functional change of variables through the Volterra integral equation
\begin{equation}
    y (\tau) = x (\tau) + \int_0^\tau \! f(x (s), s) \, ds, \hspace{1 cm} 0 \leq \tau \leq t,
\end{equation}
one can write the eq. (17) as
\begin{equation}
    \dot{y} (\tau) = \dot{\phi} (\tau).
\end{equation}
The Jacobian of this transformation can be evaluated by the discrete-time approximation
%~\vspace{2cm}
\begin{equation}
\begin{vmatrix}
1+f' (x_1,\varepsilon)\frac{\varepsilon}{2} & 0 & ... & ... & 0\\
f' (x_1,\varepsilon) & 1+f^\prime (x_2,2\varepsilon)\frac{\varepsilon}{2} & 0 & ... & 0\\
\vdots & ... & \ddots & ... & \vdots\\
\vdots & ... & ... &
\ddots & \vdots\\
f' (x_1,\varepsilon) & f^\prime (x_2,2\varepsilon) & ... & ...& 1+f^\prime (x_N,N\varepsilon)\frac{\varepsilon}{2}
\end{vmatrix},
\end{equation}
where $f^\prime \equiv \partial f/\partial x$ and $\varepsilon = t/N$. The determinant (21) becomes
\begin{equation}
    J (\varepsilon) = \prod_{n=1}^{N} \left[1 + \frac{\varepsilon f^\prime (x_n, n\varepsilon)}{2}\right].
\end{equation}
In the continuum limit, the determinant takes the form
\begin{eqnarray}
    J  &=& \lim_{\varepsilon \to 0} \exp{\left[\frac{1}{2}\sum_{n=1}^{N} \varepsilon f^\prime (x_n, n\varepsilon)\right]}\nonumber\\
     &=& \exp\left[\frac{1}{2} \int_0^t \! f^\prime(x (s), s) \, ds\right].
\end{eqnarray}
Now we can write the transition probability of the stochastic process defined by the Langevin equation in the path integral form
\begin{equation}
 W (x_t,t| x_0,0) = \int_{C(x_0,0;x_t,t)} \! \prod_{\tau=0}^{t}\frac{dx (\tau)}{\sqrt{4\pi d\tau}}\exp{\left[-\frac{1}{4}  \int_0^t \! d\tau (\dot {x} + f (x (\tau),\tau))^2\right]}\\ \exp{\left[\frac{1}{2} \int_0^t \! d\tau f^\prime(x (\tau),\tau)\right]}.
\end{equation}

The obtained Wiener path integral (24) can be generalized for solving a system of coupled Langevin equations. In the next section we shall describe such a generalization.
\subsection{Wiener path integral for a system of coupled Langevin equations}

 As a prototypical example for a system of coupled Langevin equations, let us consider a system of $\emph{N}$ Brownian particles in random media which can be treated by the generalization of the Wiener path integral method depicted in the preceding section.
For the system of $\emph{N}$ Brownian particles, there are $\emph{N}$ corresponding coupled Langevin equations
\begin{equation}
    \dot{x_i} (\tau) + f_i (x(\tau),\tau) = \dot{\Phi}_i (\tau), \hspace{2cm}       i = 1,2,...,N,
\end{equation}
and equivalently, in the matrix form:
\begin{equation}
    \dot{\mathbf{x}} (\tau) + \mathbf{f}\bigl(\mathbf{x}(\tau),\tau\bigl) = \dot{\mathbf{\Phi}} (\tau).
\end{equation}
As mentioned before, the next step is the functional change of variables
\begin{equation}
    \textbf{y}(\tau)= \textbf{x}(\tau) + \int_0^\tau \! \mathbf{f}(\mathbf{x} (s), s) \, ds, \hspace{1 cm} 0 \leq \tau \leq t,
\end{equation}
leading to
\begin{equation}
    \dot{\mathbf{y}}(\tau)\ = \dot{\mathbf{\Phi}} (\tau).
\end{equation}
Similarly to the case of (19), the Jacobian of transformation can be calculated by the discrete-time approximation,
%\vspace{2}
\begin{equation}
J(\varepsilon) =
\begin{vmatrix}
\textbf{A}(\varepsilon) & 0 & ... & ... & 0\\
* & \textbf{A}(2\varepsilon) & 0 & ... & 0\\
\vdots & ... & \ddots & ... & \vdots\\
\vdots & ... & ... & \ddots & \vdots\\
* & * & ... & * & \textbf{A}(N\varepsilon)
\end{vmatrix},
\end{equation}
where
\begin{equation}
\textbf{A} (n\varepsilon) =
\begin{pmatrix}
1+\frac{1}{2}\frac{\partial f_1 (\mathbf{x},\varepsilon)}{\partial x_1} n\varepsilon & \frac{1}{2}\frac{\partial f_1 (\mathbf{x},\varepsilon)}{\partial x_2} n\varepsilon & ... & ... & \frac{1}{2}\frac{\partial f_1 (\mathbf{x},\varepsilon)}{\partial x_N} n\varepsilon\\
\frac{1}{2}\frac{\partial f_2 (\mathbf{x},\varepsilon)}{\partial x_1} n\varepsilon & 1 + \frac{1}{2}\frac{\partial f_2 (\mathbf{x},\varepsilon)}{\partial x_2} n\varepsilon & ... & ... & ...\\
\vdots & ... & \ddots & ... & \vdots\\
\vdots & ... & ... &
\ddots & \vdots\\
\frac{1}{2}\frac{\partial f_N (\mathbf{x},\varepsilon)}{\partial x_1} n\varepsilon & \frac{1}{2}\frac{\partial f_N (\mathbf{x},\varepsilon)}{\partial x_2} n\varepsilon & ...& ... & 1 + \frac{1}{2}\frac{\partial f_N (\mathbf{x},\varepsilon)}{\partial x_N} n\varepsilon
\end{pmatrix},
\end{equation}
and the stars "*" in eq. (29) denote the matrix blocks which do not contribute to the determinant. The Jacobian of the transformation (27) is given by
\begin{equation}
    J = \lim_{\varepsilon \to 0} J (\varepsilon) = \exp\left[\frac{1}{2} \sum_{i=1}^{N} \int_0^t \! ds \frac{\partial{ f_i} (\textbf{x} (s), s)}{\partial{x_i}}\right].
\end{equation}
Now we can write the joint probability density in terms of the Wiener path integral as:
\begin{eqnarray}
 W (\textbf{x}_t,t| \textbf{x}_0,0) &=& \int_{C(\textbf{x}_0,0;\textbf{x}_t,t)} \! \prod_{\tau=0}^{t}\frac{dx_1 (\tau)}{\sqrt{4\pi d\tau}}... \prod_{\tau=0}^{t}\frac{dx_N (\tau)}{\sqrt{4\pi d\tau}}\nonumber\\
 &&
 \times \exp\left[-\frac{1}{4} \sum_{i=1}^{N} \int_0^t \! d\tau (\dot {x}_i + f_i (\textbf{x} (\tau),\tau))^2\right]\nonumber\\
 && \times \exp\left[\frac{1}{2} \sum_{i=1}^{N} \int_0^t \! d\tau \frac{\partial{f_i} (\textbf{x}(\tau),\tau)}{\partial{x_i}}\right].
\end{eqnarray}
At this level, we have developed the Wiener path integral approach for solving the system of Langevin equations and we can apply this method for the case of aggregation equations. However, there is some conceptual point concerning the meaning of the variables in the aggregation equations, namely that they are not the coordinates, but are dimensionless velocity differences, as mentioned before. So the suggested model for the aggregation of particles includes the inertia of particles, but after linearizing the equations of motion one can get the first order differential equations in the velocity space.

Now, it is straightforward to apply the above method to the case of two coupled Langevin equations. For the Langevin equations (13) describing the aggregation of particles, we can write the exact solution for the probability density in terms of path integral as follow:
\begin{eqnarray}
 W (\textbf{x}_t,t| \textbf{x}_0,0) &=& \int_{C(\textbf{x}_0,0;\textbf{x}_t,t)} \! \prod_{\tau=0}^{t}\frac{dx_1 (\tau)}{\sqrt{4\pi d\tau}}\prod_{\tau=0}^{t}\frac{dx_2 (\tau)}{\sqrt{4\pi d\tau}}\nonumber\\
 &&
 \times \exp\left[-\frac{1}{4}  \int_0^t \! d\tau \left((\dot {x}_1 + f_1)^2 + (\dot {x}_2 + f_2)^2\right)\right]\nonumber\\
 && \times  \exp\left[\frac{1}{2} \int_0^t \! d\tau \left(\frac{\partial{f_1}} {\partial{x_1}} + \frac{\partial{f_2}} {\partial{x_2}}\right)\right],
\end{eqnarray}
where
\begin{eqnarray}
f_1 &=& x_1 - \epsilon (\Gamma x_2^2 - x_1^2),\nonumber\\
f_2 &=& x_2 - x_2^{-1} + 2\epsilon x_1x_2.
\end{eqnarray}

The obtained joint probability density represents an exact solution for the system of two coupled Langevin equations for the aggregation of inertial particles. Thus, we can write the maximal Lyapunov exponent in the form of a path integral, using the eqs. (15) and (33)-(34):
\begin{eqnarray}
  \lambda_1 &=& \lim_{t\to\infty} \gamma\epsilon \, \langle x_1 \rangle = \lim_{t\to\infty} \gamma\epsilon \int_{-\infty}^{+\infty}\int_{-\infty}^{+\infty} \, dx_{1t}\, dx_{2t}\hspace{0.2cm}  \! W (\textbf{x}_t,t| \textbf{x}_0,0) \hspace{0.2cm} x_{1t}.
\end{eqnarray}
It can be seen from eq. (34) that $f_2$ and its derivative in the expressions for $W$ and $\lambda_1$ contain singularities at $x_2 = 0$. However, the region $x_2 \approx 0$ does not contribute to the probability density $W$ and the Lyapunov exponent $\lambda_1$ and overall the path integral is convergent.

The case of negative $\lambda_1$ leads to the aggregation phenomenon.
Having presented an exact solution of two coupled Langevin equations in terms of the Wiener path integral as in eq. (33), one can study all the features and properties such as Lyapunov exponent with the help of approximation methods, both theoretical and numerical, e.g. perturbation expansion in small parameters, by now very well developed lattice calculations and the saddle-point approximation.
In this way, we can also determine the points of aggregation and non-aggregation phase transition and their nature, by investigating the exact expression used to determine the Lyapunov exponent given by eq. (35). By studying the whole integrand in the exponent in eq. (35) with eq. (33), as a function of $x_1$ and $x_2$, we can find for which values of $x_1$, negative or positive, the path integral $W (\textbf{x}_t,t| \textbf{x}_0,0)$ is larger -- the region in which negative $x_1$ values dominate (as a function of $\epsilon$ and $\Gamma$) gives the region where aggregation occurs.

We will follow in this line to find analytic solutions for Lyapunov exponent by using some approximation methods.

\subsection{Evaluation of Wiener path integral for probability density in one-dimensional case: Instanton approximation}
 As an example, we can apply the above results for the analysis of aggregation phenomena in one-dimensional fluid. In the following we summarize the formulation and the results of a specific model which has been introduced by M. Wilkinson and B. Mehlig \cite{ref:1.12} for the one-dimensional fluid in a similar way as described for the three-dimensional fluid in section II.

Equations of motion for any independent particle with position $x(t)$ and momenta $p(t)$ are
\begin{equation}
\dot{x} = \frac{p}{m} , \hspace{1cm} \dot{p} = -\gamma p + f(x,t),
\end{equation}
where $\gamma$ characterizes the strength of the viscous damping \cite{ref:1.12}. The random force $f(x,t)$ is translational invariant in both space and time and the statistical properties of the force is given by the following expressions,

\begin{equation}
\langle f(x,t) \rangle = 0, \hspace{1cm} \langle f(x,t) f(x^\prime, t^\prime)\rangle = \alpha^2 \exp{\left[-\frac{\Delta x^2}{2\xi^2}\right]} \exp{\left[-\frac{\Delta t^2}{2\tau^2}\right]},
\end{equation}
where $\alpha$ denotes the magnitude of the force and $\xi$ and $\tau$ are correlation length and correlation time, respectively. The linearized version of the equation of motion can be derived as
\begin{equation}
    \delta \dot{x} = \frac{\delta p}{m} , \hspace{1cm} \delta \dot{p} = -\gamma \delta p + \partial_x f(x,t) \delta x,
\end{equation}
where $\delta x$ and $\delta p$ are small separations of positions and momentums between pairs of trajectories. One can write the equations of motion in terms of $X = \delta p/\delta x$ (where $X$ has a stationary distribution as $t\rightarrow \infty)$ as the following:

\begin{equation}
    \delta \dot{x} = \frac{X \delta x}{m},
\end{equation}
\begin{equation}
    \dot{X} = -\gamma X -\frac{X^2}{m} + \partial_x f(x,t).
\end{equation}

Since the separation $\delta x$ of two nearby trajectories has a lognormal distribution $\langle \ln|\delta x(t)/\delta x(0)| \rangle = \lambda t$, the distribution of $X$ remains stationary. From eq. (39) we obtain for the Lyapunov exponent $\lambda$

\begin{equation}
    \lambda = \frac{\langle X \rangle}{m}.
\end{equation}

 We start with the Wiener path integral analysis of this case from single Langevin equation (40). The probability density of particles in one-dimensional fluid, which is described by single Langevin equation (40), can be written in terms of a Wiener path integral. By inserting $f = \gamma X + X^2/m$ into the eq. (24) and after restoring the diffusion coefficient $D$, we obtain
 \begin{eqnarray}
 W (X_t,t|0,0) = \int_{C(0,0;X_t,t)} \! \prod_{\tau=0}^{t}\frac{dX (\tau)}{\sqrt{4\pi D d\tau}}&&\exp{\left[-\frac{1}{4D}  \int_0^t \! d\tau \left(\dot{X} + \gamma X + \frac{ X^2}{m}\right)^2\right]}\nonumber\\
 &&\times \exp{\left[ \frac{1}{2}\int_0^t \! d\tau \left(\gamma + \frac{2X}{m}\right)\right]},
\end{eqnarray}
with the last factor $\exp{\left[ \frac{1}{2}\int_0^t \! d\tau \left(\gamma + \frac{2X}{m}\right)\right]}$ being the Jacobian of transformation.
After some simplification, (42) takes the form
\begin{eqnarray}
 &&W (X_t,t|0,0) = \exp{\left[-\frac{\gamma X_t^2}{4D}- \frac{X_t^3}{6Dm} \right]}\int_{C(0,0;X_t,t)} \! \prod_{\tau=0}^{t}\frac{dX (\tau)}{\sqrt{4\pi D d\tau}}\nonumber\\
 &&\exp{\left[-\frac{1}{4D}  \int_0^t \! d\tau \left(\dot{X}^2 + \gamma^2X^2 + \frac{2\gamma X^3}{m} + \frac{X^4}{m^2} - \frac{4DX}{m}\right)\right]}\exp{\left[\frac{\gamma t}{2}\right]}.
\end{eqnarray}

Let us consider the expression for the action in the exponent in eq. (43).

We should mention that the normalization of the probability density is guaranteed by the Jacobian of transformation included in (43), although when using approximative methods we ought to normalize the probability function at each stage of approximation.

The next step is to derive an analytical expression for this probability density which is expressed in terms of Wiener path integral (43). The eq. (43) in the present form can not be solved exactly, so we try to use approximative methods in some cases. Let us first introduce new parameters and make the following change of variables:
\begin{equation}
\tilde \tau = \frac{\nu}{t} \tau = \gamma \tau , d = D t^3 / m^2, \tilde X = \frac{t}{m\nu} X  = \frac{1}{m\gamma} X.
\end{equation}
The action in terms of new variables takes the form

\begin{equation}
    S[\tilde X(\tilde \tau)]= -\frac{1}{2\varepsilon}S^\prime[\tilde X(\tilde \tau)] = -\frac{1}{2\varepsilon}  \int_0^\nu \! d\tilde \tau \left(\frac{1}{2}(\frac{d\tilde X}{d\tilde \tau})^2 + \frac{1}{2}\tilde X^2 + \tilde X ^3 + \frac{1}{2}\tilde X^4 - 2\varepsilon\tilde X\right),
\end{equation}
where $\varepsilon = d / \nu^3 = D / m^2 \gamma^3$.
The action in this form is suitable for WKB approximation in the limit $\varepsilon \rightarrow 0$.

Here, we briefly present the WKB approximation method for the evaluation of a path integral (a complete description of WKB approximation can be found, e.g. in \cite{ref:1.25}). In the WKB approximation, a general path integral can be written by using the Taylor expansion of the action $S[\tilde X(\tilde \tau)]$ around the classical solution $\tilde X_c (\tau)$, which can be obtained from the corresponding Euler-Lagrange equation for

\begin{equation}
    \frac {\delta S^\prime[\tilde X_c(\tilde \tau)]}{\delta \tilde X_c} = 0,
\end{equation}
where the subindex $¨c¨$ refers to the classical solution.
Using the Taylor expansion of (45) around the classical solution (46), gives:
\begin{eqnarray}
    &&\int \! D\tilde X (\tilde \tau)\exp{\left[-\frac{1}{\varepsilon} S[\tilde X(\tilde \tau)]\right]} =\sum_{\tilde X_c} \exp{\left[-\frac{1}{\varepsilon} S[\tilde X_c(\tilde \tau)]\right]}\nonumber\\
    &&\times \int \! D\eta (\tilde \tau)\exp{\left[-\frac{1}{2\varepsilon}\int\int \! d\tilde \tau_1 d\tilde \tau_2 \eta (\tilde \tau_1) \frac {\delta^2 S[\tilde X_c(\tilde \tau)]}{\delta \tilde X_c(\tilde \tau_1)\delta \tilde X_c(\tilde \tau_2)}\eta (\tilde \tau_2)-...\right]},
\end{eqnarray}
where the measures of integration and fluctuations $\eta (\tilde \tau)$ around the classical solution $\tilde X_c (\tilde \tau)$ have been defined as:
\begin{equation}
    D\tilde X (\tilde \tau) = \prod_{\tilde \tau=0}^{\nu}\frac{d\tilde X (\tilde \tau)}{\sqrt{4\pi \varepsilon d\tilde \tau}}, \hspace{0.2cm}\hspace{0.2cm} D\eta(\tilde \tau)=\prod_{\tilde \tau=0}^{\nu}\frac{d \eta (\tilde \tau)}{\sqrt{4\pi \varepsilon d\tilde \tau}},
\end{equation}

\begin{equation}
    \tilde X (\tilde \tau) = \tilde X_c (\tilde \tau) + \eta (\tilde \tau).
\end{equation}
In eq. (47), beside the contribution of classical configurations, there are additional path integrals over the fluctuations $\eta (\tilde \tau)$ around the classical configurations. In the first approximation, this path integral is a Gaussian one and can be calculated exactly by discretization, i.e.
\begin{eqnarray}
    &&\int \! D\eta (\tilde \tau)\exp{\left[-\frac{1}{2\varepsilon}\int\int \! d\tilde \tau_1 d\tilde \tau_2 \eta (\tilde \tau_1) \frac {\delta^2 S[\tilde X_c(\tilde \tau)]}{\delta \tilde X_c(\tilde \tau_1)\delta \tilde X_c(\tilde \tau_2)}\eta (\tilde \tau_2)\right]}\sim\nonumber\\
    && \left[ det \left(\frac {\delta^2 S[\tilde X(\tilde \tau)]}{\delta \tilde X^2}\right)_{\tilde X=\tilde X_c}\right]^{-\frac{1}{2}}.
\end{eqnarray}
This path integral is only a function of time (since $\eta(0)=\eta(\nu)=0$) which can be included in the overall normalization factor.

Let us return to the original path integral expression (43) and use the WKB approximation for its evaluation. As discussed above, in WKB approximation method one has to solve the Euler-Lagrange equation of motion to find the classical solutions. The classical equation of motion from (46) with the action in eq. (45) is :
\begin{equation}
    \ddot{\tilde X_c}-\tilde X_c-3\tilde X_c^2-2\tilde X_c^3+2\varepsilon = 0,
\end{equation}
with dot denoting the derivative with respect to $\tilde \tau$.

Instantons were introduced in the context of the turbulence for the first time by V. Gurarie and A. Migdal \cite{ref:1.26}, through the studies of the WKB approximation for the velocity distribution function of forced Burgers equation. Similarly, in our problem the path integral can provide a framework to study the instanton contribution to the aggregation phenomena, although the physical meaning of the instanton solution in this case is not yet fully clear.

Now, let us introduce a shifted variable $Z = \tilde X + \frac{1}{2}$ so that the action in eq. (45) changes its form and the new action and Lagrangian become

\begin{equation}
      S^\prime[Z] = \int_0^\nu \! d\tilde \tau \left(\frac{1}{2} \dot{Z}^2 + \frac{1}{2} (Z^2 - \frac{1}{4})^2 - 2 \varepsilon Z\right),
\end{equation}
\begin{equation}
     L = \frac{1}{2} \dot{Z}^2 + \frac{1}{2} (Z^2 - \frac{1}{4})^2 -2 \varepsilon Z.
\end{equation}
Accordingly, the Euler-Lagrange equation for general $\varepsilon$ is
\begin{equation}
    \ddot{Z}+ \frac{1}{2} Z-2Z^3+ 2 \varepsilon = 0.
\end{equation}
Eq. (54) in the limit $\varepsilon \rightarrow 0$ has the following instanton solution \cite{ref:1.27}:
\begin{equation}
    Z = \frac{1}{2} \tanh \left({\frac{1}{2}(\tilde \tau- \tilde \tau_0)}\right),
\end{equation}

The Lagrangian for this instanton solution can be calculated and the result is
\begin{equation}
    L_{inst} = \frac{1}{16}\left[ 1 - \tanh^2 \left({\frac{1}{2}(\tilde \tau- \tilde \tau_0)}\right)\right]^2,
\end{equation}
which after insertion in (52) gives the action
\begin{equation}
    S^\prime =\left. \left[ \frac{1}{12} \tanh \left({\frac{1}{2}(\tilde \tau- \tilde \tau_0)}\right) + \frac{1}{24} \cosh ^{-2}\left({\frac{1}{2}(\tilde \tau- \tilde \tau_0)}\right)\tanh \left({\frac{1}{2}(\tilde \tau- \tilde \tau_0)}\right) \right]\right|_0^\nu.
\end{equation}
The action (57) can be written in terms of $Z_\nu$ ($Z_\nu$ denotes the value of $Z$ at the time $\tilde \tau = \nu$) as
\begin{equation}
    S^\prime = \frac{1}{6} Z_\nu + \frac{1}{3} Z_\nu \dot{Z_\nu} = \frac{1}{4} Z_\nu - \frac{1}{3} Z_\nu^3,
\end{equation}
using the fact that $\dot{Z} = \frac{1}{4} - Z^2$ for (55).
By restoring the variable $\tilde X$ and the original variable $X$, we can rewrite the action (45) as
\begin{equation}
    -\frac{1}{2\varepsilon}S^\prime[\tilde X_c(\tilde \tau)] = \frac{1}{6\varepsilon} \tilde X_\nu^3 + \frac{1}{4\varepsilon} \tilde X_\nu^2,
\end{equation}
or equivalently,
\begin{equation}
    -\frac{1}{2\varepsilon}S^\prime[X_c(\tau)] = \frac{1}{6Dm} X_t^3 + \frac{\gamma}{4D} X_t^2.
\end{equation}

The contribution of this instanton solution to the probability density in the WKB approximation (47) with (43) cancels the exponential in front of the path integral (43) and leads to a zero value for the Lyapunov exponent $\lambda$ defined in eq. (41).

Besides the contribution of instanton solution, also anti-instanton solution which is another solution of eq. (54) can contribute to the evaluation of the path integral. The anti-instanton solution has the following form:
\begin{equation}
    Z = -\frac{1}{2} \tanh \left({\frac{1}{2}(\tilde \tau- \tilde \tau_0^\prime)}\right).
\end{equation}

The action can be obtained in a similar way as for the instanton solution and the result is:
\begin{equation}
    -\frac{1}{2\varepsilon}S^\prime[X_c(\tau)] = -\frac{1}{6Dm} X_t^3 - \frac{\gamma}{4D} X_t^2.
\end{equation}

Thus, using the eqs. (47) and (50) the WKB approximation for the probability density (43) yields
\begin{equation}
    W (X_t) = \beta \exp{\left[-\frac{\gamma X_t^2}{2D}- \frac{X_t^3}{3Dm} \right]}.
\end{equation}
From the definition of $\varepsilon$, it can be seen that in the limit $\varepsilon \rightarrow 0$, where the above solution is obtained, the dimensionless viscosity $\nu$ is much larger than dimensionless diffusion coefficient $d$. In this regime we can consider the Gaussian part as the main part and keeping the leading terms in the expansion of the the remaining part, we obtain
\begin{equation}
    W (X_t) = \beta \left( 1- \frac{X_t^3}{3Dm} \right) \exp{\left[-\frac{\gamma X_t^2}{2D}\right]},
\end{equation}
where $\beta$ is a normalization constant.
Hence the Lyapunov exponent can be obtained as
\begin{equation}
    \lambda = \frac{1}{m} \int_{-\infty}^{+\infty} \, dX_t\hspace{0.2cm} W (X_t) X_t = -\frac{D}{m^2\gamma^2} .
\end{equation}
We see that in the limit $\varepsilon \rightarrow 0$, the Lyapunov exponent $\lambda$ is negative so the aggregation of particles (coalescence of paths) occurs in this regime.

The obtained result $\lambda=-D/(m^2\gamma^2)$ in (65) is identical to the one previously obtained in \cite{ref:1.12} using a different approximation method.

Along this line, one can search for corrections to the existing results by expansion around the instanton solution. Also the consideration of the multi-instantons configurations will be of interest.

In the following we will continue our analysis for the evaluation of the Lyapunov exponent in the non-aggregation phase and its derivation from the path integral formalism.

Aggregation and non-aggregation phases can occur in many different cases for many different values of diffusion and viscosity coefficients, but we restrict ourselves to special cases, where the path integral is exactly solvable or is appropriate for approximative methods.
Thus, let us consider the case of a very large $D$, large $\gamma$ with $\gamma/D$ small but $\gamma^2/D$ fixed, when a non-aggregation phase can be imagined, since the effect of diffusive force is more important than the effect of viscous force. Hence, in this case the third and fourth terms , i.e. $2\gamma X^3/(Dm)$ and $X^4/(Dm^2)$ in the action in (43), can be neglected compared with the other terms. This brings the path integral expression (43) for the probability density into an exactly solvable form:

\begin{eqnarray}
 W (X_t,t|0,0) =\exp{\left[-\frac{\gamma X_t^2}{4D}\right]} \int_{C(0,0;X_t,t)} \! \prod_{\tau=0}^{t}\frac{dX (\tau)}{\sqrt{4\pi D d\tau}}&&\exp{\left[-\frac{1}{4D}  \int_0^t \! d\tau \left(\dot{X}^2 + \gamma^2X^2 - \frac{4DX}{m}\right)\right]}\nonumber\\
 &&
 \times  \exp{\left[\frac{\gamma t}{2}\right]}.
\end{eqnarray}

The expression (66) includes a Gaussian path integral, which has an exact analytical solution \cite{ref:1.25}. Here we describe the steps of calculations very briefly. Completing the quadratic form in the exponent by the change of variable $X- 2D/(m\gamma^2) = Y$, one obtains

\begin{eqnarray}
 W (X_t,t|0,0) =\exp{\left[-\frac{\gamma X_t^2}{4D}\right]} \int_{C(0,0;Y_t,t)} \! \prod_{\tau=0}^{t}\frac{dY (\tau)}{\sqrt{4\pi D d\tau}}&&\exp{\left[-\frac{1}{4D}  \int_0^t \! d\tau \left((\dot{Y})^2 + \gamma^2 Y^2 \right)\right]}\nonumber\\
 &&
 \times \exp{\left[\frac{D}{m^2\gamma^2}t\right]} \exp{\left[\frac{\gamma t}{2}\right]}.
\end{eqnarray}
After evaluation of the Gaussian path integral and restoring the original variable $X$, we reach an analytic expression for the probability density

\begin{eqnarray}
 W (X_t,t|0,0) =\left[\frac{\gamma}{4\pi D \sinh {(\gamma t)}} \right]^\frac{1}{2}&&\exp{\left[-\frac{\gamma}{4D}(X_t - \frac{2D}{m\gamma^2})^2 \coth{(\gamma t)} \right]}\nonumber\\
 &&\times \exp{\left[-\frac{\gamma X_t^2}{4D}\right]}\exp{\left[\frac{D}{m^2\gamma^2}t\right]}\exp{\left[\frac{\gamma t}{2}\right]}.
\end{eqnarray}
After taking the limit $t\rightarrow \infty$ and normalizing the probability density $W (X_t,t|0,0)$ in (68) to unity, we obtain
\begin{equation}
 W (X_t,t|0,0) =\left[\frac{\gamma}{2\pi D} \right]^\frac{1}{2}\exp{\left[-\frac{\gamma}{2D}(X_t - \frac{D}{m\gamma^2})^2\right]}.
\end{equation}

Using (69) for the probability density we find the Lyapunov exponent in this case:
\begin{equation}
    \lambda = \frac{\langle X_t \rangle}{m} = \frac{D}{m^2\gamma^2}>0.
\end{equation}
Therefore, a positive quantity for Lyapunov exponent in this case for the non-aggregation phase is obtained.

As a final remark, we should mention that the probability density $W(X_t,t|0,0)$ in (42) satisfies the Fokker-Planck equation
\begin{equation}
    \frac{\partial W (X_t,t|0,0)}{\partial t} = \frac{\partial\left[(\frac{X^2}{m}+\gamma X) W (X_t,t|0,0)\right]}{\partial X}+D\frac{\partial^2 W (X_t,t|0,0)}{\partial X^2}.
\end{equation}
The Fokker-Planck eq. (71) can be derived by writing down the expression for infinitesimally shifted $W(X_{t+\epsilon},t+\epsilon|0,0)$ as in (42) and making expansion around $X_t$ and $t$. Eq. (71) is identical to the Fokker-Planck equation used in \cite{ref:1.12}.
\subsection*{Acknowledgments}
We are much grateful to A. A. Beilinson  for several illuminating discussions. We also thank O. V. Pavlovsky for clarifying discussions on the subsection III C. The projects nos. 121720 and 127626 of the Academy of Finland are greatly acknowledged.

\end{document}